\documentclass[11pt,a4paper,fleqn]{article}
\usepackage[intlimits,sumlimits,namelimits]{amsmath}

\usepackage{epsfig}
\usepackage{amsthm}
\usepackage{amsmath}
\usepackage{amssymb}

\usepackage{braket}
\usepackage{enumerate}

\usepackage{hyperref}
\usepackage{url}

\usepackage{xcolor}
\usepackage[displaymath, mathlines]{lineno}

\usepackage{natbib}
\begin{document}

\title{\bfseries Wigner's Friend Paradox Revisited}
\author{Peter Reichert\footnote{retired from Eawag D{\"u}bendorf and ETH Z{\"u}rich, Switzerland, see \href{https://peterreichert.github.io}{https://peterreichert.github.io} for updated information; Email: \href{mailto:peter.reichert@emeriti.eawag.ch}{peter.reichert@emeriti.eawag.ch}, \href{mailto:peter.reichert@usys.ethz.ch}{peter.reichert@usys.ethz.ch}, ORCID: \href{https://orcid.org/0000-0001-7832-4257}{0000-0001-7832-4257}} \hspace{0.1ex} and Markus Enz\footnote{Email: \href{mailto:enzmar@student.ethz.ch}{enzmar@student.ethz.ch}, ORCID: \href{https://orcid.org/0009-0004-6589-7944}{0009-0004-6589-7944}}}
\maketitle

\begin{minipage}{0.9\textwidth}
By assuming (i) a universal, observer-independent quantum state, by considering (ii) measurement processes with wave function collapse as part of quantum mechanical time evolution (also within isolated systems), and by (iii) clearly distinguishing the state of a quantum system from the knowledge of conscious observers about this state, we suggest a modified Copenhagen interpretation of quantum mechanics that resolves Wigner's Friend and extended Wigner's Friend paradoxes.
The suggested interpretation leads to differences in some outcome probabilities compared to previous analyses which, in principle, make it possible to falsify the suggested modifications or the previous analyses (or both).
The fundamental problem of the incompleteness of quantum mechanics in the Copenhagen interpretation regarding the lack of a precise definition and a mechanistic description of the measurement process (including the collapse of the wave function) is not addressed in this study.
However, the argumentation regarding the resolution of Wigner's Friend paradoxes also applies to attempts for such a mechanistic description using collapse models with stochastic extensions to the Schr{\"o}dinger equation or trying to model wave function collapse with unitary time evolution and irreversibility of the measurement process resulting from quantum statistical mechanics.
\end{minipage}

\clearpage

\section{Introduction}
\label{cha:Intro}

Wigner's Friend paradox \citep{Wigner_1961_MindBodyQuestion} consists of apparently contradictory states of a quantum system identified by two different observers.
We suggest resolving this paradox, as well as that of extended Wigner's Friend paradoxes (e.g.\ \citet{FrauchigerRenner_2018_ConsistencyQM}), by assuming a universal, observer-independent quantum state, by considering measurement processes with wave function collapse as part of quantum mechanical time evolution also within isolated quantum systems, and by clearly distinguishing the (objective) state description of a quantum system from the (intersubjective) knowledge of conscious observers about this state.\\

This paper is only about modifying the Copenhagen interpretation of quantum mechanics \citep{Bohr_1935_CopenhagenInterpretation,Heisenberg_1958_PhysicsAndPhilosophy,Omnes_1992_CopenhagenInterpretation,Faye_2024_CopenhagenInterpretation}
to resolve Wigner's Friend and related paradoxes while still relying on seemingly reasonable assumptions.
We still rely on the axiom of wave function collapse related to measurement processes as a pragmatic approach to make the theory applicable as long as no final solution for a mechanistic theory to these processes is found.
The fundamental problem of this incompleteness of quantum mechanics (regarding the lack of a precise definition and a mechanistic description of the measurement process and the collapse of the wave function) is not addressed in this paper.

\section{Assumptions}
\label{cha:Assumptions}

We suggest modifications to typical versions of the Copenhagen interpretation of quantum mechanics
\citep{Bohr_1935_CopenhagenInterpretation,Heisenberg_1958_PhysicsAndPhilosophy,Omnes_1992_CopenhagenInterpretation,Faye_2024_CopenhagenInterpretation}
by making the following modifying assumptions: 
\begin{enumerate}
\item The state of a quantum system is fully described by a normalized vector in a Hilbert space or by a wave function (vectors that differ only by a constant, complex factor of modulus unity represent the same state). The state vector is universal and thus independent of observers.
\item The time evolution of the quantum system is given by the continuous, deterministic, unitary evolution according to the Schr{\"o}dinger equation and by measurement processes that lead to projections of the state vector to the eigenspace of the observed eigenvalue of an observable which is realized randomly according to the Born probabilities.
``Observed'' means here a registration of the outcome in a macroscopic (quantum) system, irrespective of whether a conscious observer reads this result.
Also isolated (quantum) systems underlay both kinds of time evolution processes.
\item Conscious observers may have perfect, partial, or missing information about the state of a quantum system. Partial information can often be expressed as a probability distribution across states.
\end{enumerate}
The universality of the state vector is a modification to some versions of the Copenhagen interpretation in which it is dependent on the observer.
However, the major deviating point from most previous versions is the inclusion of measurement processes as part of the quantum mechanical time evolution (also within isolated systems) in addition to unitary evolution.
We suggest accepting measurement processes as part of quantum mechanical time evolution for mainly two reasons:
First, looking at the whole universe, measurement processes with their intrinsic stochasticity obviously take place without an external, ``classical'' observer.
Second, many attempts to a mechanistic description of the measurement process, e.g.\ stochastic modifications to the Schr{\"o}dinger equation \citep[and many more studies]{Bassi_2013_ModelsOfWaveFunctionCollapse}, the ETH approach \citep{Froehlich_2024_CompletionQM}, and attempts to explain measurement processes by unitary evolution and quantum statistical mechanics \citep{Allahverdyan_2013_UnderstandingQuantumMeasurement} would lead to random evolution and wave function collapse without the need for an external observer.
In most of these attempts, the basis of the measurement process consists of the interaction of the observed quantum system with a macroscopic (``large'') quantum system which leads to a macroscopic registration of the result by a ``pointer'' \citep{Naus_2021_QuantumMechanicalMeasurementProcess} or mean magnetization \citep{Allahverdyan_2003_CurieWeissModel} that can (optionally) be read by a conscious observer.
The conscious observer, however, is not needed in the suggested interpretation for the measurement process including the wave function collapse.
As the measurement process is not yet fully understood, for this analysis of the Wigner's Friend paradox, we just keep measurement with wave function collapse as an axiom in the hope that this can be replaced by a mechanistic theory in the future, leaving the analysis of the Wigner's Friend paradox valid.\\

In addition to the registration of the measurement result by the macroscopic, quantum mechanical ``measurement device'', we intend to describe the knowledge of conscious observers about the state of the quantum system.
When having this in mind, it is essential to use an appropriate mathematical framework to describe knowledge.
We describe epistemic uncertainty, in our case the partial knowledge or beliefs of conscious observers about a quantum mechanical state, by probability distributions across possible states.
The most important reasons to do so are
\begin{itemize}
\item
the consistency when beliefs are operationalized by asking individuals for indifference between lotteries \citep{Howson_1989_ScientificReasoning};
\item
the formulation of conditional beliefs that fulfill ``reasonable'' requirements \citep{Cox_1946_Probability,Lindely_1982_InevitabilityOfProbability};
\item
the consistency of beliefs with underlying objective probabilities of random events after the random event has ocurred but the result is still unknown \citep{Reichert_2015_EnvironmentalDecisionSupport}.
\end{itemize}
There are alternative theories to mathematically represent beliefs \citep{Colyvan_2008_IsProbabilityTheOnlyCoherentApproachToUncertainty,Helton_2004_RepresentationsOfEpistemicUncertainty} that are, in our view, conceptually less satisfying as they do not fulfill the consistency conditions listed above.
To represent knowledge in a scientific context, it is relevant to use intersubjective probabilities \citep{Gillies_1991_IntersubjectiveProbability,Gillies_2000_PhilosophicalTheoriesOfProbability}, i.e.\ probabilities about which different experts agree (or can be assumed to agree based on a justification that relies on established principles).
This is not a new concept, as the scientific process, e.g.\ represented by the peer review process, is based on the intersubjective interpretation of (objective) scientific results.
In the context of the quantum mechanical measurement process, it is easy to follow this approach.
Whenever the Born probabilities are known, different scientists would agree about their beliefs of a measurement outcome once it has ocurred (no randomness involved any more) but they do not yet know the result (the objective probabilities before the random event become intersubjective probabilities after the event has occurred and the result has not yet been seen).
However, it is very important to distinguish between the measurement process (interaction between a system and a ``measurement device'' resulting in macroscopic registration of the result and wave function collapse) and the update of the knowledge of a conscious, external observer about the result.
In the Copenhagen interpretation of quantum mechanics, the first process is a random quantum mechanical event accompanied by the collapse of the wave function, whereas the update of the knowledge of the external observer is typically not described as a quantum mechanical process.
However, it is in principle possible to include the conscious observer as part of a quantum system having in mind that this would lead to an overwhelmingly complex quantum system that cannot be treated without strong simplifications.

\section{Wigner's Friend}
\label{cha:Wigner}

We assume time evolution according to the Schr{\"o}dinger equation with a Hamiltonian equals to zero, except for the interaction with the measurement devices during the short times of measurement processes, and collapses of the wave function as parts of measurement processes.
We use the notation as used by \citet{FrauchigerRenner_2018_ConsistencyQM}.\\

In Wigner's Friend paradox \citep[and many newer descriptions]{Wigner_1961_MindBodyQuestion} it is assumed that a friend, F, is taking a spin measurement along the $z$-axis in an isolated lab and that Wigner, W, later-on measures the state of the lab as an external observer.
Both agents are assumed to know the initial state of the spin and the lab and the procedural steps of the experimental setup.\\

The procedure, the state, and the knowledge of the observers, F and W, are now described step by step based on the modified Copenhagen assumptions listed in section \ref{cha:Assumptions}.
These steps, as well as the states, the results registered by the measurement devices, and the knowledge of conscious observers who read the registered results when they become available, are also summarized in Table \ref{tab:WignersFriend}.\\

The initial state of the spin is assumed to be given by
\begin{equation}
   \ket{\mathrm{ini}}_\mathrm{S} 
   = \ket{\rightarrow}_\mathrm{S}
   = \sqrt{\frac{1}{2}} \Bigl( \ket{\uparrow}_\mathrm{S} + \ket{\downarrow}_\mathrm{S} \Bigr)
   \quad .
\label{equ:Wigner_spin_ini}
\end{equation}
The lab is assumed to consist of the spin, S, a measurement device, D, and the (conscious) friend, F, which leads to a state space
\begin{equation}
   \mathrm{L} = \mathrm{S} \otimes \mathrm{D} \otimes \mathrm{F} \quad .
\label{equ:Wigner_statespace}
\end{equation}
The initial state of the lab (assuming the device pointer points to zero and F knows the initial state of the spin; the results should be independent of the initial states of D and F) is given by
\begin{multline}
   \ket{\mathrm{ini}}_\mathrm{L} 
   = \ket{\rightarrow}_\mathrm{S} \; \otimes \; \ket{"z= 0"}_\mathrm{D} \; \otimes \; \ket{"\psi_\mathrm{S} = \ket{\rightarrow}"}_\mathrm{F} \\
   = \sqrt{\frac{1}{2}} \Bigl( \ket{\uparrow}_\mathrm{S} + \ket{\downarrow}_\mathrm{S} \Bigr) \; \otimes \; \ket{"z=0"}_\mathrm{D} \; \otimes \; \ket{"\psi_\mathrm{S} = \ket{\rightarrow}"}_\mathrm{F}
   \quad .
\label{equ:Wigner_lab_ini}
\end{multline}
Here, $z$ is the position of the pointer of the measurement device, and $\psi_\mathrm{S}$ is the knowledge of F about the state of the spin gained from the knowledge of the initial state of the spin.
This state is described by the row labeled "step 0" in Table \ref{tab:WignersFriend}. 
At that time, there is no registered result by devices in the lab nor externally (see columns labelled ``R'') and both, F and W know the state with certainty (probability 1, listed in the columns labelled ``K'').\\

After the measurement of the spin along the $z$-axis, due to the collapse of its wave function, the final state of the spin is either
\begin{subequations}
\begin{equation}
   \ket{\uparrow}_\mathrm{S}
\end{equation}
or
\begin{equation}
   \ket{\downarrow}_\mathrm{S}
   \quad .
\end{equation}
\end{subequations}
The state of the device pointer corresponds to the resulting spin and the same is true for the knowledge of F after reading the device pointer.
This leads the state of the lab which is either
\begin{subequations}
\begin{equation}
   \ket{\scriptstyle + \frac{1}{2}}_\mathrm{L}
   = \ket{\uparrow}_\mathrm{S} \; \otimes \; \ket{"z={\scriptstyle + \frac{1}{2} }"}_\mathrm{D} \; \otimes \; \ket{"\psi_\mathrm{S} = \ket{\uparrow}"}_\mathrm{F}
\label{equ:Wigner_lab_spinup}
\end{equation}
or
\begin{equation}
   \ket{\scriptstyle - \frac{1}{2}}_\mathrm{L}
   = \ket{\downarrow}_\mathrm{S} \; \otimes \; \ket{"z={\scriptstyle - \frac{1}{2} }"}_\mathrm{D} \; \otimes \; \ket{"\psi_\mathrm{S} = \ket{\downarrow}"}_\mathrm{F}
\label{equ:Wigner_lab_spindown}
\end{equation}
\label{equ:Wigner_lab}
\end{subequations}
\hspace{-1.8ex}
depending on the measurement outcome.
This state is described by the row labeled "step 1" in Table \ref{tab:WignersFriend}. 
The registered result by the device in the lab is either $+\frac{1}{2}$ or $-\frac{1}{2}$ (column labelled ``R$_\mathrm{D}$'') and there is no registered result for the external observer, W (column labelled ``R$_\mathrm{W}$'').
Consequently, the knowledge of F about the result is certainty (probability 1) for the registered outcome (column labelled ``K$_\mathrm{F}$'') whereas the knowledge of W is given by a 50\% probability for the two outcomes of $+\frac{1}{2}$ or $-\frac{1}{2}$ (column labelled ``K$_\mathrm{F}$'') because W knows the setup of the experiment and the reasoning of W is according to the rules of quantum mechanics.
For this reason, the objective Born probabilities before the measurement took place, are taken as intersubjective, epistemic probabilities for the result after the measurement has taken place but the result is not (yet) known to W\footnote{Note that this probability distribution represents more knowledge than knowledge described by the density matrix $\frac{1}{2} \ket{\uparrow} \bra{\uparrow}$ +$\frac{1}{2} \ket{\downarrow} \bra{\downarrow}$ because it is known that the state is a mixture of the states $\ket{\uparrow}$ and $\ket{\downarrow}$ (simplified notation) whereas the density matrix would allow mixtures of other states as it is equal to (e.g.) $\frac{1}{2} \frac{\ket{\uparrow}+\ket{\downarrow}}{\sqrt{2}} \frac{\bra{\uparrow}+\bra{\downarrow}}{\sqrt{2}}$ + $\frac{1}{2} \frac{\ket{\uparrow}-\ket{\downarrow}}{\sqrt{2}} \frac{\bra{\uparrow}-\bra{\downarrow}}{\sqrt{2}}$. The higher information content of the probability distribution over the density matrix may not be of practical relevance as the difference does not result in different measurement outcomes.}.\\

\begin{table}[htb]
\small
\begin{center}
\begin{tabular}{c|c|c|c|c|c|c}
step & true state & eq. & \multicolumn{4}{|l}{R: registered result} \\
&&& \multicolumn{4}{|l}{K: probab.\ knowl.\ of obs.} \\
&&&  \multicolumn{2}{|c}{L} & \multicolumn{2}{|c}{ext.\ world} \\
&&& \hspace{0.5ex}R$_\mathrm{D}$\hspace{0.5ex} & \hspace{0.5ex}K$_\mathrm{F}$\hspace{0.5ex} & \hspace{0.5ex}R$_\mathrm{W}$\hspace{0.5ex} & \hspace{0.5ex}K$_\mathrm{W}$\hspace{0.5ex} \\
\hline
0 & $\ket{\mathrm{ini}}_\mathrm{L} = \sqrt{\frac{1}{2}} \Bigl( \ket{\uparrow}_\mathrm{S} + \ket{\downarrow}_\mathrm{S} \Bigr) \; \otimes \; \ket{"z=0"}_\mathrm{D} \; \otimes \; \ket{"\psi_\mathrm{S} = \ket{\rightarrow}"}_\mathrm{F}$ & \ref{equ:Wigner_lab_ini} & - & 1 & - & 1 \\
\hline
1 & $\ket{\scriptstyle + \frac{1}{2}}_\mathrm{L} = \ket{\uparrow}_\mathrm{S} \; \otimes \; \ket{"z={\scriptstyle + \frac{1}{2} }"}_\mathrm{D} \; \otimes \; \ket{"\psi_\mathrm{S} = \ket{\uparrow}"}_\mathrm{F}$ & \ref{equ:Wigner_lab_spinup} & $+\frac{1}{2}$ & 1 & - & $\frac{1}{2}$ \\
   & or &&&&& \\
   & $\ket{\scriptstyle - \frac{1}{2}}_\mathrm{L} = \ket{\downarrow}_\mathrm{S} \; \otimes \; \ket{"z={\scriptstyle - \frac{1}{2} }"}_\mathrm{D} \; \otimes \; \ket{"\psi_\mathrm{S} = \ket{\downarrow}"}_\mathrm{F}$ & \ref{equ:Wigner_lab_spindown} & $-\frac{1}{2}$ & 1 & - & $\frac{1}{2}$ \\
\hline
2 & $\ket{\scriptstyle + \frac{1}{2}}_\mathrm{L} = \ket{\uparrow}_\mathrm{S} \; \otimes \; \ket{"z={\scriptstyle + \frac{1}{2} }"}_\mathrm{D} \; \otimes \; \ket{"\psi_\mathrm{S} = \ket{\uparrow}"}_\mathrm{F}$ & \ref{equ:Wigner_lab_spinup} & $+\frac{1}{2}$ & 1 & $+\frac{1}{2}$ & 1 \\
   & or &&&&& \\
   & $\ket{\scriptstyle - \frac{1}{2}}_\mathrm{L} = \ket{\downarrow}_\mathrm{S} \; \otimes \; \ket{"z={\scriptstyle - \frac{1}{2} }"}_\mathrm{D} \; \otimes \; \ket{"\psi_\mathrm{S} = \ket{\downarrow}"}_\mathrm{F}$ & \ref{equ:Wigner_lab_spindown} & $-\frac{1}{2}$ & 1 & $-\frac{1}{2}$ & 1 \\
\end{tabular}
\end{center}
\caption{Summary of the true state and the knowledge of F and W during the experiment.
Step 0: initial state known to F and W. Step 1: collapse of the wave function after measurement by F: F knows the result, W only knows the measurement setup and the state before measurement and thus the probabilities of the possible results. Step 2: measurement by W would lead to another collapse, but as the measurement is along the same axis, the state is already an eigenstate and does not change; only the knowledge of W is updated.}
\label{tab:WignersFriend}
\end{table}

If now W performs a measurement, in principle, this would change the state and involve a wave function collapse.
However, as the measurement is again along the $z$-axis of the spin, this ``collapse'' does not change the wave function as it is already in the correct eigenstate regarding the spin, the device and F due to the measurement by F.
This state is described by the row labeled "step 2" in Table \ref{tab:WignersFriend}. 
This row is identical to the row labelled "step 1" regarding the registered result and the knowledge of F, but now there is the identical registered result for W and W has become certain about the result by reading the registered result of the measurement device.\\

The usual assumption in the Wigner's Friend setup is that, due to the interaction of the spin with the measurement device, the state would turn into a linear combination of the two states \eqref{equ:Wigner_lab_spinup} and \eqref{equ:Wigner_lab_spindown}, but, in contrast to the situation for F, for the external observer, W, there would not be a collapse into one of these states
These different states of the same system as assessed by F and W and the different collapse time points are the core of Wigner's Friend paradox.\\

As with our modified assumptions the state is assumed to be observer-independent and we track the knowledge of conscious observers about this state, at no point in time there is an inconsistency about the state of the system or the assessment of the state by the observers; there is just a time interval, in which W only has partial knowledge about the state.
This is not a specific feature of quantum mechanics; only the specific state description and the Born probabilities are specific to quantum mechanics.
That observers can have full, partial or no information about the state of a system is a trivial fact across whichever theories of nature.

\section{Extended Wigner's Friend Setup}
\label{cha:ExtendedWigner}

\citet{FrauchigerRenner_2018_ConsistencyQM} suggested an extended Wigner's Friend setup and demonstrated the possibility of inconsistent results by two observers that follow the rules of quantum mechanics.
This paper was heavily discussed in the literature \citep[and many more references cited therein]{Lazarovici_2019_ConsistencyQM,Polychronakos_2024_ConsistencyQM,delRio_2024_ConsistencyQM}.\\

The extended Wigner's Friend setup, as summarized by \citet{Lazarovici_2019_ConsistencyQM} is given as follows:
Two quantum experiments are performed by experimentalists $\overline{\mathrm{F}}$ and F in their respective labs $\overline{\mathrm{L}}$ and L,
while the other two are “Wigner’s friend–type” measurements that outside observers $\overline{\mathrm{W}}$ and W perform on the
labs, which are assumed to have been perfectly isolated up to this point.
The experiment starts with a quantum coin toss: $\overline{\mathrm{F}}$ measures a qubit (``coin-toss'') and prepares a spin-1/2 particle depending on the outcome.
If the result of the coin-toss is heads, she prepares the spin-state $\ket{\downarrow}_{\overline{\mathrm{S}}}$, if the result is tails, she prepares the spin-state $\ket{\rightarrow}_{\overline{\mathrm{S}}}$ (the bar over S indicates that the spin is in the lab $\overline{\mathrm{L}}$). 
F then receives the so prepared particle and performs a spin-measurement in $z$-direction, obtaining a result of $z$ = +1/2 or $z$ = $-$1/2.
Next, $\overline{\mathrm{W}}$ performs a quantum measurement on the entire lab $\overline{\mathrm{L}}$ based on a basis (see Table 2 in \citet{FrauchigerRenner_2018_ConsistencyQM})
\begin{subequations}
\begin{equation}
   \ket{\overline{\mathrm{ok}}}_{\overline{\mathrm{L}}} 
   = \sqrt{\frac{1}{2}} \Bigl( \ket{\mathrm{h}}_{\overline{\mathrm{L}}} - \ket{\mathrm{t}}_{\overline{\mathrm{L}}} \Bigr)
   \quad , \quad
   \ket{\overline{\mathrm{fail}}}_{\overline{\mathrm{L}}} 
   = \sqrt{\frac{1}{2}} \Bigl( \ket{\mathrm{h}}_{\overline{\mathrm{L}}} + \ket{\mathrm{t}}_{\overline{\mathrm{L}}} \Bigr)
   \quad ,
\label{equ:okbar}
\end{equation}
where $\ket{\mathrm{h}}_{\overline{\mathrm{L}}}$ represents the state of the lab $\overline{\mathrm{L}}$ if the result of the coin toss was equal to heads and the spin has been prepared in the state $\ket{\downarrow}_{\overline{\mathrm{S}}}$ and $\ket{\mathrm{t}}_{\overline{\mathrm{L}}}$  represents the state of the lab $\overline{\mathrm{L}}$ if the result of the coin toss was equal to heads and the spin has been prepared in the state $\ket{\rightarrow}_{\overline{\mathrm{S}}}$ (see equation \eqref{equ:WignersFriendExt_Coin} below).
Finally, W performs a quantum measurement on the entire lab L based on a basis (see Table 2 in \citet{FrauchigerRenner_2018_ConsistencyQM})
\begin{equation}
   \ket{\mathrm{ok}}_\mathrm{L} 
   = \sqrt{\frac{1}{2}} \Bigl( \ket{\scriptstyle - \frac{1}{2}}_{\mathrm{L}} - \ket{\scriptstyle + \frac{1}{2}}_{\mathrm{L}} \Bigr)
   \quad , \quad
   \ket{\mathrm{fail}}_\mathrm{L} 
   = \sqrt{\frac{1}{2}} \Bigl( \ket{\scriptstyle - \frac{1}{2}}_{\mathrm{L}} + \ket{\scriptstyle + \frac{1}{2}}_{\mathrm{L}} \Bigr)
\label{equ:ok}
\end{equation}
\end{subequations}
with $\ket{\scriptstyle - \frac{1}{2}}_{\mathrm{L}}$ and $\ket{\scriptstyle + \frac{1}{2}}_{\mathrm{L}}$ according to equation \eqref{equ:Wigner_lab}.
The procedure is repeated until $\overline{\mathrm{W}}$ and W obtain the outcomes $\ket{\overline{\mathrm{ok}}}_{\overline{\mathrm{L}}}$ and $\ket{\mathrm{ok}}_\mathrm{L}$.\\

The state spaces of the two labs are given by
\begin{equation}
   \mathrm{\overline{L}} = \mathrm{\overline{R}} \otimes \overline{\mathrm{D}} \otimes \overline{\mathrm{S}} \otimes \overline{\mathrm{F}}
   \quad \mbox{and} \quad
   \mathrm{L} = \mathrm{S} \otimes \mathrm{D} \otimes \mathrm{F} \quad ,
\label{equ:WignersFriendExt_statespace}
\end{equation}
where $\mathrm{\overline{R}}$ represents the (random) coin toss, $\overline{\mathrm{D}}$ the measurement device for the coin toss, $\overline{\mathrm{S}}$ the spin in lab $\overline{\mathrm{L}}$, $\overline{\mathrm{F}}$ the friend in lab $\overline{\mathrm{L}}$, and S, D and F are the same as in the basic Wigner's Friend setup described in the text before equation \eqref{equ:Wigner_statespace} (note that $\overline{\mathrm{L}}$ describes the state of the lab before the spin is transferred whereas L describes the lab after the transfer).\\

The procedure, the state, the registered results by the measurement devices, and the knowledge of the observers, $\overline{\mathrm{F}}$, F, $\overline{\mathrm{W}}$ and W, are now described step by step based on the modified Copenhagen assumptions listed in section \ref{cha:Assumptions}.
These steps are also summarized in Table \ref{tab:WignersFriendExt}.
For ease of notation, we only show the relevant parts of the overall state vector; nevertheless, we always have a joint quantum state across both laboratories, but treat $\overline{\mathrm{W}}$ and W as external observers.
Note that despite we describe the knowledge of conscious observers $\overline{\mathrm{F}}$, F, $\overline{\mathrm{W}}$ and W, these are not required for the experimental setup.
The spin in the lab $\overline{\mathrm{L}}$ can be initialized by a machine that does it conditional on the registered result of the coin toss and the spin can then be transferred to the lab L and be measured in this lab.
Analogously, the subsequent measurements on the whole labs $\overline{\mathrm{L}}$ and L based on bases \eqref{equ:okbar} and \eqref{equ:ok}, respectively, can be initiated automatically and the results be registered (note that without the observers $\overline{\mathrm{F}}$ and F the state spaces of the labs reduce from $\mathrm{\overline{R}} \otimes \overline{\mathrm{D}} \otimes \overline{\mathrm{S}} \otimes \overline{\mathrm{F}}$ to $\mathrm{\overline{R}} \otimes \overline{\mathrm{D}} \otimes \overline{\mathrm{S}}$ and from $\mathrm{S} \otimes \mathrm{D} \otimes \mathrm{F}$ to $\mathrm{S} \otimes \mathrm{D}$, respectively, and the last factors in the equations \eqref{equ:WignersFriendExt_Coin} and \eqref{equ:Wigner_lab} drop).
This procedure can also be repeated automatically unless the outcomes are $\ket{\overline{\mathrm{ok}}}_{\overline{\mathrm{L}}}$ and $\ket{\mathrm{ok}}_\mathrm{L}$.
Nevertheless, the description of the knowledge of conscious observers is important to demonstrate that there is also no inconsistency involved for them.\\

The initial state of the quantum coin is given by (see Table 1 in \citet{FrauchigerRenner_2018_ConsistencyQM}):
\begin{equation}
   \ket{\mathrm{ini}}_{\overline{\mathrm{R}}} 
   = \sqrt{\frac{1}{3}} \ket{\mathrm{h}}_{\overline{\mathrm{R}}} + \sqrt{\frac{2}{3}} \ket{\mathrm{t}}_{\overline{\mathrm{R}}}
   \quad .
\label{equ:WignersFriendExt_Coin_Ini}
\end{equation}
This state is described by the row labeled "step 0" in Table \ref{tab:WignersFriendExt}. 
At that time, there are no registered results by devices in the labs nor externally (see columns labelled ``R'') and all observers,  $\overline{\mathrm{F}}$, F,  $\overline{\mathrm{W}}$ and W know the state with certainty (probability 1, listed in the columns labelled ``K'')).\\

After measuring the coin toss and preparing the spin according to the instruction given above, the state of the lab $\overline{\mathrm{L}}$ is either given by
\begin{subequations}
\begin{equation}
   \ket{\mathrm{h}}_{\overline{\mathrm{L}}} = \ket{\mathrm{h}}_{\overline{\mathrm{R}}} \otimes \; \ket{"\mathrm{coin=h}"}_{\overline{\mathrm{D}}} \otimes \ket{\downarrow}_{\overline{\mathrm{S}}} \otimes \; \ket{"\psi_{\overline{\mathrm{RS}}} = \ket{\mathrm{h}} \otimes \ket{\downarrow}"}_{\overline{\mathrm{F}}}
\label{equ:WignersFriendExt_Coin_h}
\end{equation}
or by
\begin{multline}
   \ket{\mathrm{t}}_{\overline{\mathrm{L}}} = \ket{\mathrm{t}}_{\overline{\mathrm{R}}} \otimes \; \ket{"\mathrm{coin=t}"}_{\overline{\mathrm{D}}} \otimes \ket{\rightarrow}_{\overline{\mathrm{S}}} \otimes \; \ket{"\psi_{\overline{\mathrm{RS}}} = \ket{\mathrm{t}} \otimes \ket{\rightarrow}"}_{\overline{\mathrm{F}}} \\
   = \ket{\mathrm{t}}_{\overline{\mathrm{R}}} \otimes \; \ket{"\mathrm{coin=t}"}_{\overline{\mathrm{D}}} \otimes \sqrt{\frac{1}{2}} \Bigl( \ket{\uparrow}_{\overline{\mathrm{S}}} + \ket{\downarrow}_{\overline{\mathrm{S}}} \Bigr) \otimes \; \ket{"\psi_{\overline{\mathrm{RS}}} = \ket{\mathrm{t}} \otimes \sqrt{\frac{1}{2}} \Bigl( \ket{\uparrow} + \ket{\downarrow} \Bigr)"}_{\overline{\mathrm{F}}}
\label{equ:WignersFriendExt_Coin_t}
\end{multline}
\label{equ:WignersFriendExt_Coin}
\end{subequations}
depending on the measurement outcome.
The probability of getting the state according to equation \eqref{equ:WignersFriendExt_Coin_h} is 1/3, that of getting the state according to equation \eqref{equ:WignersFriendExt_Coin_t} is 2/3.
As $\overline{\mathrm{F}}$ reads the result of the coin toss and prepared the spin, she has perfect knowledge of this state.
In contrast, the observers F, $\overline{\mathrm{W}}$ and W only know that the measurement has taken place without knowing the result.
However, as they know the initial state \eqref{equ:WignersFriendExt_Coin_Ini} and the rules of quantum mechanics, they can accept these probabilities as their intersubjective knowledge about the state (see step 1a in Table \ref{tab:WignersFriendExt}).
In particular, due to the wave function collapse induced by the measurement, they know that the state is no longer equal to the initial state given by equation \eqref{equ:WignersFriendExt_Coin_Ini} nor to a linear combination of the states \eqref{equ:WignersFriendExt_Coin_h} and \eqref{equ:WignersFriendExt_Coin_t} that had developed through the interaction of the spin with the measurement device.
This is a deviation from the assumptions of \citet{FrauchigerRenner_2018_ConsistencyQM} resulting from our modified assumptions of the time evolution of a (isolated) system in which a measurement takes place (see assumptions formulated in section \ref{cha:Assumptions}).\\

The spin is now transferred to the lab L and its state is given by either
\begin{subequations}
\begin{equation}
   \ket{\downarrow}_{\mathrm{S}}
\label{equ:WignersFriendExt_Coin_h_spinL}
\end{equation}
or
\begin{equation}
   \ket{\rightarrow}_{\mathrm{S}}
   = \sqrt{\frac{1}{2}} \Bigl( \ket{\uparrow}_{\mathrm{S}} + \ket{\downarrow}_{\mathrm{S}} \Bigr)
\label{equ:WignersFriendExt_Coin_t_spinL}
\end{equation}
\end{subequations}
(note that $\ket{\downarrow}_{\overline{\mathrm{S}}}$ became $\ket{\downarrow}_{\mathrm{S}}$ and $\ket{\rightarrow}_{\overline{\mathrm{S}}}$ became $\ket{\rightarrow}_{\mathrm{S}}$ to indicate that the spin has been transferred from the lab $\overline{\mathrm{L}}$ to the lab L).
Still, $\overline{\mathrm{F}}$ has perfect knowledge about the state, whereas the other observers still only have intersubjective, probabilistic knowledge about the state (see step 1b in Table \ref{tab:WignersFriendExt}).\\

After the measurement of the spin along the $z$-axis by F, the spin and the lab end up in one of the states (see also equation \ref{equ:Wigner_lab} of the original Wigner's Friend paradox)
\begin{subequations}
\begin{equation}
   \ket{\downarrow}_{\mathrm{S}}
   \quad , \quad
   \ket{\scriptstyle - \frac{1}{2}}_{\mathrm{L}}
   = \ket{\downarrow}_\mathrm{S} \; \otimes \; \ket{"z={\scriptstyle - \frac{1}{2} }"}_\mathrm{D} \; \otimes \; \ket{"\psi_\mathrm{S} = \ket{\downarrow}"}_\mathrm{F}
\label{equ:WignersFriendExt_spindown}
\end{equation}
or
\begin{equation}
   \ket{\uparrow}_{\mathrm{S}}
   \quad , \quad
   \ket{\scriptstyle + \frac{1}{2}}_{\mathrm{L}}
   = \ket{\uparrow}_\mathrm{S} \; \otimes \; \ket{"z={\scriptstyle + \frac{1}{2} }"}_\mathrm{D} \; \otimes \; \ket{"\psi_\mathrm{S} = \ket{\uparrow}"}_\mathrm{F} \quad .
\label{equ:WignersFriendExt_spinup}
\end{equation}
\end{subequations}
Which result is realized is known to F from reading the registered result.
However, as $\overline{\mathrm{F}}$ prepared the spin but does not know the measurement result, she only knows the state for sure if the preparation was made in the state $\ket{\downarrow}_{\overline{\mathrm{S}}}$ (result of the coin toss was heads). 
Otherwise (result of the coin toss was tails) her knowledge is just that the probability is 50\% for each of the outcomes $ \ket{\downarrow}_\mathrm{S}$ and $ \ket{\uparrow}_\mathrm{S}$ (see step 2a in Table \ref{tab:WignersFriendExt}).
As the external observers, $\overline{\mathrm{W}}$ and W, do not know the outcome of the coin toss, there intersubjective probabilities for the two outcomes are 2/3 and 1/3 as can be derived from the combined state of the two labs which can be
\begin{subequations}
\begin{equation}
   \ket{\mathrm{h}}_{\overline{\mathrm{R}}} \otimes \ket{\scriptstyle - \frac{1}{2}}_{\mathrm{L}}
\label{equ:WignersFriendExt_L_h}
\end{equation}
or
\begin{equation}
   \ket{\mathrm{t}}_{\overline{\mathrm{R}}} \otimes \ket{\scriptstyle - \frac{1}{2}}_{\mathrm{L}}
\label{equ:WignersFriendExt_L_t_down}
\end{equation}
or
\begin{equation}
   \ket{\mathrm{t}}_{\overline{\mathrm{R}}} \otimes \ket{\scriptstyle + \frac{1}{2}}_{\mathrm{L}}
   \quad
\label{equ:WignersFriendExt_L_t_up}
\end{equation}
\label{equ:WignersFriendExt_L}
\end{subequations}
each with a probability of 1/3 (see step 2b in Table \ref{tab:WignersFriendExt}).\\
\begin{table}[h!]
\small
\begin{center}
\begin{tabular}{c|c|c|c|c|c|c|c|c|c|c}
step & true state & eq. & \multicolumn{8}{|l}{R: registered result by the measurement device} \\
& (only most relevant parts && \multicolumn{8}{|l}{K: knowledge of conscious observer (probability of state)} \\
& of the full state vector) &&  \multicolumn{2}{|c}{$\overline{\mathrm{L}}$} & \multicolumn{2}{|c}{L} & \multicolumn{4}{|c}{external world} \\
&&&  R$_{\overline{\mathrm{D}}}$ & K$_{\overline{\mathrm{F}}}$ & R$_\mathrm{D}$ & K$_\mathrm{F}$ & R$_{\overline{\mathrm{W}}}$ & K$_{\overline{\mathrm{W}}}$ & R$_\mathrm{W}$ & K$_\mathrm{W}$ \\
\hline
0 & $\ket{\mathrm{ini}}_{\overline{\mathrm{R}}} = \sqrt{\frac{1}{3}} \ket{\mathrm{h}}_{\overline{\mathrm{R}}} + \sqrt{\frac{2}{3}} \ket{\mathrm{t}}_{\overline{\mathrm{R}}}$ & \ref{equ:WignersFriendExt_Coin_Ini} & - & 1 & - & 1 & - & 1 & - & 1 \\
\hline
1a & $\ket{\mathrm{h}}_{\overline{\mathrm{R}}} \otimes \ket{\downarrow}_{\overline{\mathrm{S}}}$ & \ref{equ:WignersFriendExt_Coin_h} & h & 1 & - & $\frac{1}{3}$ & - & $\frac{1}{3}$ & - & $\frac{1}{3}$ \\
   & or &&&&&&&& \\
   & $\ket{\mathrm{t}}_{\overline{\mathrm{R}}} \otimes \ket{\rightarrow}_{\overline{\mathrm{S}}} = \ket{\mathrm{t}}_{\overline{\mathrm{R}}} \otimes \sqrt{\frac{1}{2}} \Bigl( \ket{\uparrow}_{\overline{\mathrm{S}}} + \ket{\downarrow}_{\overline{\mathrm{S}}} \Bigr)$ & \ref{equ:WignersFriendExt_Coin_t} & t & 1 & - & $\frac{2}{3}$ & - & $\frac{2}{3}$ & - & $\frac{2}{3}$ \\
\hline
1b & $\ket{\downarrow}_{\mathrm{S}}$ & \ref{equ:WignersFriendExt_Coin_h_spinL} & h & 1 & - & $\frac{1}{3}$ & - & $\frac{1}{3}$ & - & $\frac{1}{3}$ \\
   & or &&&&&&&& \\
   & $\ket{\rightarrow}_{\mathrm{S}} =\sqrt{\frac{1}{2}} \Bigl( \ket{\uparrow}_{\mathrm{S}} + \ket{\downarrow}_{\mathrm{S}} \Bigr)$ & \ref{equ:WignersFriendExt_Coin_t_spinL} & t & 1 & - & $\frac{2}{3}$ & - & $\frac{2}{3}$ & - & $\frac{2}{3}$ \\
\hline
2a & $\ket{\scriptstyle - \frac{1}{2}}_{\mathrm{L}}$ & \ref{equ:WignersFriendExt_spindown} & h or t & h:1, t:$\frac{1}{2}$ & $-\frac{1}{2}$ & 1 & - & $\frac{2}{3}$ & - & $\frac{2}{3}$ \\
   & or &&&&&&&& \\
   & $\ket{\scriptstyle + \frac{1}{2}}_{\mathrm{L}}$ & \ref{equ:WignersFriendExt_spinup} & t & $\frac{1}{2}$ & $+\frac{1}{2}$ & 1 & - & $\frac{1}{3}$ & - & $\frac{1}{3}$ \\
\hline
2b & $\ket{\mathrm{h}}_{\overline{\mathrm{R}}} \otimes \ket{\scriptstyle - \frac{1}{2}}_{\mathrm{L}}$ & \ref{equ:WignersFriendExt_L_h} & h & 1 & $-\frac{1}{2}$ & $\frac{1}{2}$ & - & $\frac{1}{3}$ & - & $\frac{1}{3}$ \\
   & or &&&&&&&& \\
   & $\ket{\mathrm{t}}_{\overline{\mathrm{R}}} \otimes \ket{\scriptstyle - \frac{1}{2}}_{\mathrm{L}}$ & \ref{equ:WignersFriendExt_L_t_down} & t & $\frac{1}{2}$ & $-\frac{1}{2}$ & $\frac{1}{2}$ & - & $\frac{1}{3}$ & - & $\frac{1}{3}$ \\
   & or &&&&&&&& \\
   & $\ket{\mathrm{t}}_{\overline{\mathrm{R}}} \otimes \ket{\scriptstyle + \frac{1}{2}}_{\mathrm{L}}$ & \ref{equ:WignersFriendExt_L_t_up} & t & $\frac{1}{2}$ & $+\frac{1}{2}$ & 1 & - & $\frac{1}{3}$ & - & $\frac{1}{3}$ \\
\hline
3 & $\ket{\overline{\mathrm{ok}}}_{\overline{\mathrm{L}}} \otimes \ket{\mathrm{ok}}_\mathrm{L}$ & \ref{equ:okbar},\ref{equ:ok} & $\overline{\mathrm{ok}}^{\,*}$ & $\frac{1}{2}{\,^{**}}$ & ok$^{\,*}$ & $\frac{1}{2}{\,^{**}}$ & $\overline{\mathrm{ok}}$ & $\frac{1}{2}{\,^{**}}$ & ok & $\frac{1}{2}{\,^{**}}$ \\
   & or & &&&&&&&& \\
   & $\ket{\overline{\mathrm{fail}}}_{\overline{\mathrm{L}}} \otimes \ket{\mathrm{ok}}_\mathrm{L}$ & \ref{equ:okbar},\ref{equ:ok} & $\overline{\mathrm{fail}}^{\,*}$ & $\frac{1}{2}{\,^{**}}$ & ok$^{\,*}$ & $\frac{1}{2}{\,^{**}}$ & $\overline{\mathrm{fail}}$ & $\frac{1}{2}{\,^{**}}$ & ok & $\frac{1}{2}{\,^{**}}$ \\
   & or &&&&&&&&& \\
   & $\ket{\overline{\mathrm{ok}}}_{\overline{\mathrm{L}}} \otimes \ket{\mathrm{fail}}_\mathrm{L}$ & \ref{equ:okbar},\ref{equ:ok} & $\overline{\mathrm{ok}}^{\,*}$ & $\frac{1}{2}{\,^{**}}$ & fail$^{\,*}$ & $\frac{1}{2}{\,^{**}}$ & $\overline{\mathrm{ok}}$ & $\frac{1}{2}{\,^{**}}$ & fail & $\frac{1}{2}{\,^{**}}$ \\
   & or & &&&&&&&& \\
   & $\ket{\overline{\mathrm{fail}}}_{\overline{\mathrm{L}}} \otimes \ket{\mathrm{fail}}_\mathrm{L}$ & \ref{equ:okbar},\ref{equ:ok} & $\overline{\mathrm{fail}}^{\,*}$ & $\frac{1}{2}{\,^{**}}$ & fail$^{\,*}$ & $\frac{1}{2}{\,^{**}}$ & $\overline{\mathrm{fail}}$ & $\frac{1}{2}{\,^{**}}$ & fail & $\frac{1}{2}{\,^{**}}$ \\
\end{tabular}
\end{center}
\caption{Summary of the true state and the knowledge of $\overline{\mathrm{F}}$, F, $\overline{\mathrm{W}}$ and W during the experiment.
Step 0: initial state known to all. 
Step 1: observation of the outcome of the coin toss, collapse of the wave function, preparation and transfer of the spin (a: state of both labs. b: state of lab L only).
Step 2: measurement of the spin by F (a: state of lab L. b: state of both labs).
Step 3: after measurements by $\overline{\mathrm{W}}$ and W (irrespective of order).\\
$^*\,$ These results refer only to the states of the coin registered by the device $\overline{\mathrm{D}}$ in lab $\overline{\mathrm{L}}$ and of the spin registered by the device D in lab L, not to the total states of the labs.\\
$^{**}$ Knowledge is perfect about the state in the corresponding lab; conditional on this result, the probability is 50\% for the combination with the possible results in the other lab.}
\label{tab:WignersFriendExt}
\end{table}

Finally, according to the setup of the experiment, the external observers $\overline{\mathrm{W}}$ and W perform their measurements on the labs $\overline{\mathrm{L}}$ and L based on the bases \eqref{equ:okbar} and \eqref{equ:ok}, respectively.
Depending on the outcomes of the measurements, this leads to the collapse of the wave function of the lab $\overline{\mathrm{L}}$ to either $ \ket{\overline{\mathrm{ok}}}_{\overline{\mathrm{L}}}$ or $ \ket{\overline{\mathrm{fail}}}_{\overline{\mathrm{L}}}$ \eqref{equ:okbar} and of the lab L to either $ \ket{\mathrm{ok}}_{\mathrm{L}}$ or $ \ket{\mathrm{fail}}_{\mathrm{L}}$ \eqref{equ:ok}.
After reading the result, the outcome in the corresponding lab is known to $\overline{\mathrm{W}}$ and W, but as they do not know the result of the measurement of the other lab, their knowledge of the full state consists of 50\% probability for a combination of the known state with the two options of the state for the other lab (see step 3 in Table \ref{tab:WignersFriendExt}).
As the measurements of the external observers, $\overline{\mathrm{W}}$ and W, are regarding the whole labs, $\overline{\mathrm{L}}$ and L, the collapses of the wave functions of the labs to one of the four states listed in step 3 in Table \ref{tab:WignersFriendExt} (see definitions of the bases in the equations \ref{equ:okbar} and \ref{equ:ok}), these collapses also update the measurement devices and the knowledge of the observers (who are quantum mechanical systems) about the spins in their labs.
This leads again to the knowledge of these observers with a probability of 50\% for a combination of the known state with the two options of the state for the other lab (see step 3 in Table \ref{tab:WignersFriendExt}).
If the external observers would only measure the coin and the spin instead of the whole labs, the wave functions of the devices and the internal observers would not collapse and their knowledge would be 25\% for each of the four states in step 3 of Table \ref{tab:WignersFriendExt}.\\

The probability of getting the outcomes $\ket{\overline{\mathrm{ok}}}_{\overline{\mathrm{L}}}$ and $\ket{\mathrm{ok}}_\mathrm{L}$ can be calculated as follows:
{\small
\begin{multline}
   p_{\overline{\mathrm{ok}} \, \mathrm{ok}} \\
   = \frac{1}{3} \Bigl( \bra{\overline{\mathrm{ok}}}_{\overline{\mathrm{L}}} \otimes \bra{\mathrm{ok}}_\mathrm{L} \;
                              \ket{\mathrm{h}}_{\overline{\mathrm{L}}} \otimes \ket{\scriptstyle - \frac{1}{2}}_{\mathrm{L}} \Bigr)^2
   + \frac{1}{3} \Bigl( \bra{\overline{\mathrm{ok}}}_{\overline{\mathrm{L}}} \otimes \bra{\mathrm{ok}}_\mathrm{L} \;
                              \ket{\mathrm{t}}_{\overline{\mathrm{L}}} \otimes \ket{\scriptstyle - \frac{1}{2}}_{\mathrm{L}} \Bigr)^2
   + \frac{1}{3} \Bigl( \bra{\overline{\mathrm{ok}}}_{\overline{\mathrm{L}}} \otimes \bra{\mathrm{ok}}_\mathrm{L} \;
                              \ket{\mathrm{t}}_{\overline{\mathrm{L}}} \otimes \ket{\scriptstyle + \frac{1}{2}}_{\mathrm{L}} \Bigr)^2 \\
   = \frac{1}{3} \Bigl( \bra{\overline{\mathrm{ok}}}_{\overline{\mathrm{L}}} \ket{\mathrm{h}}_{\overline{\mathrm{L}}} \cdot 
                              \bra{\mathrm{ok}}_\mathrm{L} \ket{\scriptstyle - \frac{1}{2}}_{\mathrm{L}} \Bigr)^2
   + \frac{1}{3} \Bigl( \bra{\overline{\mathrm{ok}}}_{\overline{\mathrm{L}}} \ket{\mathrm{t}}_{\overline{\mathrm{L}}} \cdot
                              \bra{\mathrm{ok}}_\mathrm{L} \ket{\scriptstyle - \frac{1}{2}}_{\mathrm{L}} \Bigr)^2
   + \frac{1}{3} \Bigl( \bra{\overline{\mathrm{ok}}}_{\overline{\mathrm{L}}} \ket{\mathrm{t}}_{\overline{\mathrm{L}}} \cdot
                              \bra{\mathrm{ok}}_\mathrm{L} \ket{\scriptstyle + \frac{1}{2}}_{\mathrm{L}} \Bigr)^2 \\
   = \frac{1}{3} \Bigl( \frac{1}{2} \cdot \frac{1}{2} \Bigr) + \frac{1}{3} \Bigl( \frac{1}{2} \cdot \frac{1}{2} \Bigr) + \frac{1}{3} \Bigl( \frac{1}{2} \cdot \frac{1}{2} \Bigr)
   = \frac{1}{12} + \frac{1}{12} + \frac{1}{12}
   = \frac{1}{4} \hfill
\label{equ:WignersFriendExt_Pokok}
\end{multline}
}\\
\hspace{-1.5ex}
where the three terms in each row result from the three equations \eqref{equ:WignersFriendExt_L_h}, \eqref{equ:WignersFriendExt_L_t_down} and \eqref{equ:WignersFriendExt_L_t_up} and the factors 1/3 are coming from the probabilities of the outcomes corresponding to these equations.
Note that each of these three outcomes contribute by 1/12 to the overall probability of 3/12 = 1/4.
The essential difference of this result to the result by \citet{FrauchigerRenner_2018_ConsistencyQM} originates from replacing their result for the coin toss of tails,
\begin{equation}
   \ket{\mathrm{t}}_{\overline{\mathrm{R}}} \otimes  \sqrt{\frac{1}{2}} \Bigl( \ket{\scriptstyle - \frac{1}{2}}_{\mathrm{L}} + \ket{\scriptstyle + \frac{1}{2}}_{\mathrm{L}} \Bigr)
\label{equ:WignersFriendExt_L_t_FR}
\end{equation}
(with probability 2/3) by the two equations \eqref{equ:WignersFriendExt_L_t_down} and \eqref{equ:WignersFriendExt_L_t_up} (each with a probability of 1/3) due to the collapse of the wave function to the eigenstate of the observed eigenvalue.
The expression \eqref{equ:WignersFriendExt_L_t_FR} is orthogonal to $\ket{\mathrm{ok}}_\mathrm{L}$ according to equation \eqref{equ:ok} and does thus not contribute to the probability of $p_{\overline{\mathrm{ok}} \, \mathrm{ok}}$.
For this reason, \citet{FrauchigerRenner_2018_ConsistencyQM} get a probability of $p_{\overline{\mathrm{ok}} \, \mathrm{ok}}$ = 1/12.
This orthogonality is destroyed by the collapse leading to the equations \eqref{equ:WignersFriendExt_L_t_down} and \eqref{equ:WignersFriendExt_L_t_up} which both contribute by 1/12 to $p_{\overline{\mathrm{ok}} \, \mathrm{ok}}$ and we end up with an overall probability $p_{\overline{\mathrm{ok}} \, \mathrm{ok}}$ = 3/12 = 1/4 according to equation \eqref{equ:WignersFriendExt_Pokok}.
The probabilities for the other outcome combinations are the same:
\begin{equation}
   p_{\overline{\mathrm{ok}} \, \mathrm{fail}}
   = p_{\overline{\mathrm{fail}} \, \mathrm{ok}}
   = p_{\overline{\mathrm{fail}} \, \mathrm{fail}}
   = \frac{1}{4}
   \quad .
\end{equation}
\vspace{0.5ex}

During the whole process, we have a unique state of the (overall) system and for some observers full and for others partial information about this state, but the partial information is always consistent with the true state (the true state has always probability larger than zero for all observers).
With the modified assumptions of the Copenhagen interpretation, we thus have no inconsistency of quantum mechanics in this thought experiment.


\section{Discussion}
\label{cha:Discussion}

With some modifications to most frequently used versions of the Copenhagen interpretation of quantum mechanics, namely (see section \ref{cha:Assumptions} for a more detailed description) by 
\begin{enumerate}
\item assuming a universal state vector that is independent of observers,
\item accepting measurement processes that lead to the projection (``collapse'') of the state vector to the eigenspace of the observed eigenvalue of an observable (which is realized randomly according to the Born probabilities) in addition to continuous, deterministic, unitary evolution according to the Schr{\"o}dinger equation as part of quantum mechanical time evolution,
\item describing the state of knowledge of conscious observers about the state of a quantum system,
\end{enumerate}
we can resolve any inconsistencies resulting from Wigner's Friend paradox \citep{Wigner_1961_MindBodyQuestion} and extended Wigner's Friend setups \citep{FrauchigerRenner_2018_ConsistencyQM}.\\

\citet{FrauchigerRenner_2018_ConsistencyQM} (see also \citet{delRio_2024_ConsistencyQM}) formulated criteria (Q), (C) and (S) (see below) and proved that in none of the investigated interpretations of quantum mechanics all three of these criteria could be fulfilled jointly.
This is a problem of the consistency of quantum mechanics.
With an important modification (Q') to the requirement (Q), we argue that the suggested modifications to the Copenhagen interpretation lead to the joint fulfillment of all three criteria (Q'), (C) and (S):
\begin{itemize}
\item[(Q')]
Validity of quantum theory as used in the analysis at the relevant scales:\\
In the original formulation of the requirement (Q) a specific interpretation of quantum mechanics had to be valid; obviously, this excludes any (potentially innovative) modifications (see below for other modifications than the one suggested in this paper that lead to similar results). With the modified criterion (Q') we require the consistent application of a suggested version of quantum mechanics (rather than a specific pre-defined version). In our case, we include probabilistic measurements that follow the Born rule and are accompanied by wave function collapse as an intergral part of quantum mechanical time evolution in addition to deterministic, unitary time evolution.
\item[(C)]
Consistency among agents:\\
All agents base their reasoning on the same version of quantum mechanics and there is no disagreement between agents. However, some agents may know the result of measurements and others may only have partial information about it.
\item[(S)]
Single outcomes:\\
Measurements lead to unique results that are registered independently of observers. Observers may have complete, partial or no information about these results.
\end{itemize}
\vspace{1ex}

It is trivial that modified assumptions (and modified criteria) can lead to different conclusions.
The key point is thus to discuss whether the modifications in the assumptions make sense.
We justify the suggested modifications 1 - 3 as listed above and explained in more detail in section \ref{cha:Assumptions} as follows:
\begin{enumerate}
\item There is a an intensive, ongoing discussion about the reality of the quantum state (see e.g.\ \citet{Leifer_2014_IsTheQuantumStateReal} for an extensive review).
In accordance with most versions of the Copenhagen interpretation \citep{Leifer_2014_IsTheQuantumStateReal,Faye_2024_CopenhagenInterpretation},
\citet{FrauchigerRenner_2018_ConsistencyQM} assume different (epistemic) states of the system assessed by different observers.
In particular, they assume that the states of the isolated labs $\overline{\mathrm{L}}$ and L for the outside observers develop according to the Schr{\"o}dinger equation despite that the outside observers know that a qubit measurement (in lab $\overline{\mathrm{L}}$) and a spin measurement along the $z$-axis (in lab L) take place.
On the other hand, for the observers $\overline{\mathrm{F}}$ and F the state may change (depending on the interpretation) due to their measurements. 
With the assumption of a universal state, this source of discrepancy is eliminated.
There is significant literature that supports such a universal state \citep{Pusey_2012_RealityOfQuantumState,Ringbauer_2015_MeasurementOnRealityOfWavefunction,Brown_2019_TheRealityOfTheWavefunction} which is also assumed in spontaneous collapse models \citep{Bassi_2013_ModelsOfWaveFunctionCollapse} and in the ETH approach \citep{Froehlich_2024_CompletionQM}.
On the other hand, there is also support for observer-dependence of the quantum state \citep{Harrigan_2010_EpistemicViewOfQuantumStates,Fuchs_2014_QBism}; the most obvious advantage of these interpretations is that they solve the measurement problem as the collapse of the wave function just consists of updating the (Bayesian) probability distribution of the observers.
In summary, there are arguments in favor of a universal state vector as well as arguments in favor of an observer-dependent state vector.
Given this ambiguity, we chose the option of a universal, observer-independent state which seems to represent the simplest quantum mechanical theory and which makes it easier to avoid inconsistency problems.
\item It is the standard assumption of quantum mechanics, that the state of isolated systems develops unitarily (and deterministically) according to the Schr{\"o}dinger equation.
However, radioactive decay, photon absorption by atoms, photon emission, measurement processes, and many more observations, demonstrate the need for a stochastic description of the evolution of the state of individual quantum systems (see e.g.\ section 1.1 in \citet{Froehlich_2024_CompletionQM}).
The Schr{\"o}dinger equation may still be appropriate for the description of ensembles of identical systems.
As the cause and mechanisms of this stochasticity is currently still unknown, as a pragmatic, ``temporary'' solution (until at least one of the current trials for a more complete theory will be established), we propose to include an empirical ``measurement process'' with wave function collapse as part of quantum mechanical time evolution (axiom 2 above).
As such a measurement process is not well-defined, this is not satisfying, but it allows us to demonstrate the resolution of reported inconsistency problems based on arguments that should remain valid once a better theory will become available.
There are various attempts for such better theories.
Collapse models add empirical stochastic terms to the Schr{\"o}dinger equation \citep{Ghirardi_1986_GRWmodel,Ghirardi_1990_CSLmodel,Bassi_2013_ModelsOfWaveFunctionCollapse}, however, large ranges of the empirical parameters of these model have already been excluded experimentally, see Figure 2 in \citet{Carlesso_2022_TestsOfCollapseModels}.
There are also attempts to include the cause of the stochastic evolution of the state vector, such as gravitational fluctuations \citep{Diosi_1989_UniversalReduction,Penrose_1996_GravitysRoleInQuantumStateReduction} or interactions with the quantized electromagnetic field \citep{Froehlich_2024_CompletionQM}.
The ETH approach by \citet{Froehlich_2024_CompletionQM} is based on a more general non-linear and stochastic time evolution of individual systems that reduces to the linear and deterministic Schr{\"o}dinger equation for ensembles.
Finally, attempts to explain measurement processes by unitary evolution and quantum statistical mechanics \citep{Allahverdyan_2013_UnderstandingQuantumMeasurement} may also be an interesting direction of research to resolve the measurement problem.
All of these attempts do not need a conscious observer and our arguments for the resolution of inconsistencies of quantum mechanics in the context of extended Wigner's Friend problems apply to these theories as well.
\item In our view it is very important to clearly distinguish the state of a system and the knowledge of a conscious observer about the state.
In many contexts, knowledge or beliefs of individuals can be quantitatively described by subjective probability distributions 
\citep[see also section \ref{cha:Assumptions} for more details]{Howson_1989_ScientificReasoning,Cox_1946_Probability,Lindely_1982_InevitabilityOfProbability,Reichert_2015_EnvironmentalDecisionSupport}.
To represent knowledge in a scientific context, it is relevant to use intersubjective probabilities \citep{Gillies_1991_IntersubjectiveProbability,Gillies_2000_PhilosophicalTheoriesOfProbability}, i.e.\ probabilities about which different experts agree.
This is particularly easy in the case of quantum measurements of a known state for which we can assume that different experts would agree about the Born probabilities of the outcomes and the associated states.
Before the measurement process was initiated, these probabilities describe the random measurement process; after the measurement has taken place but the result has not yet been observed, these probabilities build a natural intersubjective probability distribution across potential outcomes and associated states that characterizes the knowledge of an ``observer'' who is not informed about the result.
This is exactly the situation of the external observer(s) in Wigner's Friend and extended Wigner's Friend settings.
When being informed about the result, the updated information will be certainty about the outcome and the associated state; when performing a new measurement, reading the newly registered result will again reflect certainty.
\end{enumerate}
Given these justifications, we believe that the modified assumptions are reasonable and they resolve the inconsistencies of Wigner's Friend and extended Wigner's Friend scenarios.\\

As our modified assumptions lead to differences in probabilities of some outcomes, they are, in principle, testable.
The nature of these differences and their test is very similar to the differences between the Copenhagen and the Many Worlds interpretations of quantum mechanics as discussed by \citet{Deutsch_1985_QuantumTheory}.
In particular, his equations (69) and (70) reflect exactly the same difference in wave functions as the equation (3) of \citet{FrauchigerRenner_2018_ConsistencyQM} and our equation \eqref{equ:Wigner_lab} or of equation \eqref{equ:WignersFriendExt_L_t_FR} and the equations \eqref{equ:WignersFriendExt_L_t_down} and \eqref{equ:WignersFriendExt_L_t_up}.
In the first equations, it is assumed that the state is not affected for the external observer despite the measurement has taken place in the isolated lab, whereas in the second equations, due to the collapse of the wave function (for all observers), the wave function changes to the eigenfunction of the observed value of the observable.
Proceeding with one or the other wave function, leads to the (in principle) observable differences in \citet{Deutsch_1985_QuantumTheory} as well as in this study.

\section{Conclusions}
\label{cha:Conclusions}

The objective of this paper was to develop a resolution of the inconsistency of quantum mechanics in an extended Wigner's Friend setting identified by \citet{FrauchigerRenner_2018_ConsistencyQM} by modifying the Copenhagen interpretation of quantum mechanics.
Our results demonstrate that this is possible with modifications to the assumptions, in particular by assuming the state of the quantum mechanical system to be universal, by allowing for continuous, unitary evolution as well as wave function collapses as parts of quantum mechanical time evolution, and by carefully distinguishing the (objective) state from the (intersubjective) knowledge of conscious observers about the state.
These assumptions do not seem to be less reasonable than assuming observer-dependent quantum mechanical states as assumed by \citet{FrauchigerRenner_2018_ConsistencyQM} and in many more studies.
Our analysis applies to the suggested version of the Copenhagen interpretation with wave function collapse as a postulate of quantum mechanics, as well as to spontaneous collapse models that are based on stochastic extensions of the Schr{\"o}dinger equation, the ETH approach, and unitary evolution with irreversibility resulting from quantum statistical mechanics.
As none of these approaches is yet sufficiently developed, we used the postulate-based approach as a pragmatic base for our argumentation.
We hope that the collapse postulate will later be replaced by a better justified theory.\\

In addition to resolving the inconsistency identified by \citet{FrauchigerRenner_2018_ConsistencyQM}, our modified assumptions lead to different probabilities of some outcomes.
In principle, this makes it possible to falsify the suggested modifications or the previous analyses (or both).\\

\section*{Acknowledgements}

Our thoughts were inspired by the Quantum Foundations Seminar at ETH Zurich led by Beate Elisabeth Asenbeck and Viktoria Kabel in the spring semester 2026.
We thank Carlo Albert and Beate Elisabeth Asenbeck for their comments to an early draft version of this paper and in particular Viktoria Kabel for her useful hints for improving this paper.

\section*{Competing interests}

The authors declare no competing interests.
Search engines and AI tools have only be used to support literature search.

\bibliography{refs}
\end{document}